\pdfoutput=1
\documentclass[12pt,nofootinbib]{iopart}

\usepackage[linktoc=page,colorlinks,urlcolor=blue,citecolor=blue,linkcolor=blue]{hyperref}

\usepackage{amssymb}
\usepackage{graphicx}
\usepackage{natbib}
\usepackage{lineno}
\usepackage{overpic}
\usepackage[font=small,labelfont=bf,
   justification=raggedright,
   format=plain]{caption}
\setcitestyle{numbers,square,citesep={,}}
\newcommand{\ket}[1]{|#1\rangle}
\newcommand{\snl}{Sandia National Laboratories, Albuquerque, New Mexico 87185, USA}
\newcommand{\cint}{Center for Integrated Nanotechnologies, Sandia National Laboratories, Albuquerque, New Mexico 87123, USA}

\newcommand{\brown}{Department of Physics, Brown University, Providence, Rhode Island 02912, USA}

\newcommand{\carbon}{$^{13}$C}
\newcommand{\nitrogen}{$^{15}$N}

\usepackage{verbatim}

\begin{document}

\title{Mitigation of Nitrogen Vacancy Ionization from Material Integration for Quantum Sensing }

\date{\today}
\author{Jacob Henshaw} \address{\cint}
\author{Pauli Kehayias} \address{\snl}
\author{Luca Basso} \address{\cint}
\author{Michael Jaris} \address{\cint}
\author{Rong Cong} \address{\brown}
\author{Michael Titze} \address{\snl}
\author{Tzu-Ming Lu} \address{\cint}
\author{Michael P. Lilly} \address{\cint}
\author{Andrew M. Mounce} \address{\cint}
\hfill
\hfill
\begin{abstract}
The nitrogen-vacancy (NV) color center in diamond has demonstrated great promise in a wide range of quantum sensing. Recently, there have been a series of proposals and experiments using NV centers to detect spin noise of quantum materials near the diamond surface. This is a rich complex area of study with novel nano-magnetism and electronic behavior, that the NV center would be ideal for sensing. However, due to the electronic properties of the NV itself and its host material, getting high quality NV centers within nanometers of such systems is challenging. Band bending caused by space charges formed at the metal-semiconductor interface force the NV center into its insensitive charge states. Here, we investigate optimizing this interface by depositing thin metal films and thin insulating layers on a series of NV ensembles at different depths to characterize the impact of metal films on different ensemble depths. We find an improvement of coherence and dephasing times we attribute to ionization of other paramagnetic defects. The insulating layer of alumina between the metal and diamond provide improved photoluminescence and higher sensitivity in all modes of sensing as compared to direct contact with the metal, providing as much as a factor of 2 increase in sensitivity, decrease of integration time by a factor of 4, for NV $T_1$ relaxometry measurements.
\end{abstract}
\maketitle
\section{Introduction}
The rapidly-developing field of 2-D materials has the opportunity to provide advances in the fields of data storage, magnetometry, and quantum information processing. However, due to their low-dimensional nature, established bulk characterization techniques, such as nuclear magnetic resonance (NMR) and electron paramagnetic resonance (EPR) spectroscopy, lack the sensitivity to properly probe electron dynamics responsible for phenomena such as magnetism and superconductivity. A series of measurements have been proposed\cite{demler_2d_SC,demler_magneticIns} and realized\cite{NV_MBT_Relaxometry,NV_SC_Screening}offering the nitrogen-vacancy (NV) center as a new probes of low dimensional electronic phases of quantum materials. Due to its long lived electronic spin state and ease of read-out and control, the NV is an excellent sensor of magnetic and electric noise with bandwidth ranging from DC to GHz provided through a variety of sensing modalities\cite{CrIMagnetometry,CrBr3_magnetometry,Henshaw_NQR,NV_MBT_Relaxometry,T1_Relaxometry_Overview}.

A challenge of working with NV centers is preserving the notable spin properties as the NVs form closer to the diamond surface. Without extensive oxidation treatments\cite{Henshaw_NQR,KMF_selective_oxidation,UVOzone}, the diamond surface can provide a strong upward band bending, depleting the NV of it electrons and converting the NV to its magnetically-insensitive neutral and positive charge states. Additionally, the surface provides a source of noise that worsens the NV's dephasing, decoherence, and relaxation\cite{depth_decoherence,surface_noise}. These problems can be exacerbated by the integration of metals, conductive materials, or materials with large work functions to the diamond surface.

The addition of a metal to the diamond surface forms a positive charge at the interface that is compensated by the negative charge in traps and defects in the diamond, like the aforementioned NV center. Other sources of negative charges, such as substitutional nitrogen ($\mathrm{N_s}$), may also be ionized in this process. This creates an involved competition with regards to sensing, where some ionization of $\mathrm{N_s}$ maybe beneficial in improving coherence properties, but too much ionization may result in destabilizing the $\mathrm{NV^-}$

Here, we explore how dense ensembles of NVs are affected by the integration of such materials. We deposited a thin film of metal onto the diamond surface and characterized the spin properties the NVs under the metal film, and compared to an uncoated area. We repeated this for a range of NV ensembles of different depths. We find that, depending on the depth of the NVs, spin properties such as $T_2^*$ and $T_2$ times, improve by as much as a factor of 1.4 and 1.6 respectively, while relaxation times and photo-luminescence rates are quenched due to proximity to the thin metal film. We tune this effect by adding a thin layer of alumina in between the metal and diamond, and find the photoluminescence intensity improves while preserving some degree of improvement to $T_2^*$ and $T_2$ times and providing minimal impact on relaxation times, preserving utility for $T_1$ relaxometry. We conclude by estimating the shot-noise-limited sensitivity in different sensing modalities, and find that for all forms of sensing, the addition of the insulating layer improves the sensitivity as compared to direct contact with the thin film.

\section{Methods}
\subsection{Sample Preparation}
\begin{figure}
\begin{overpic}{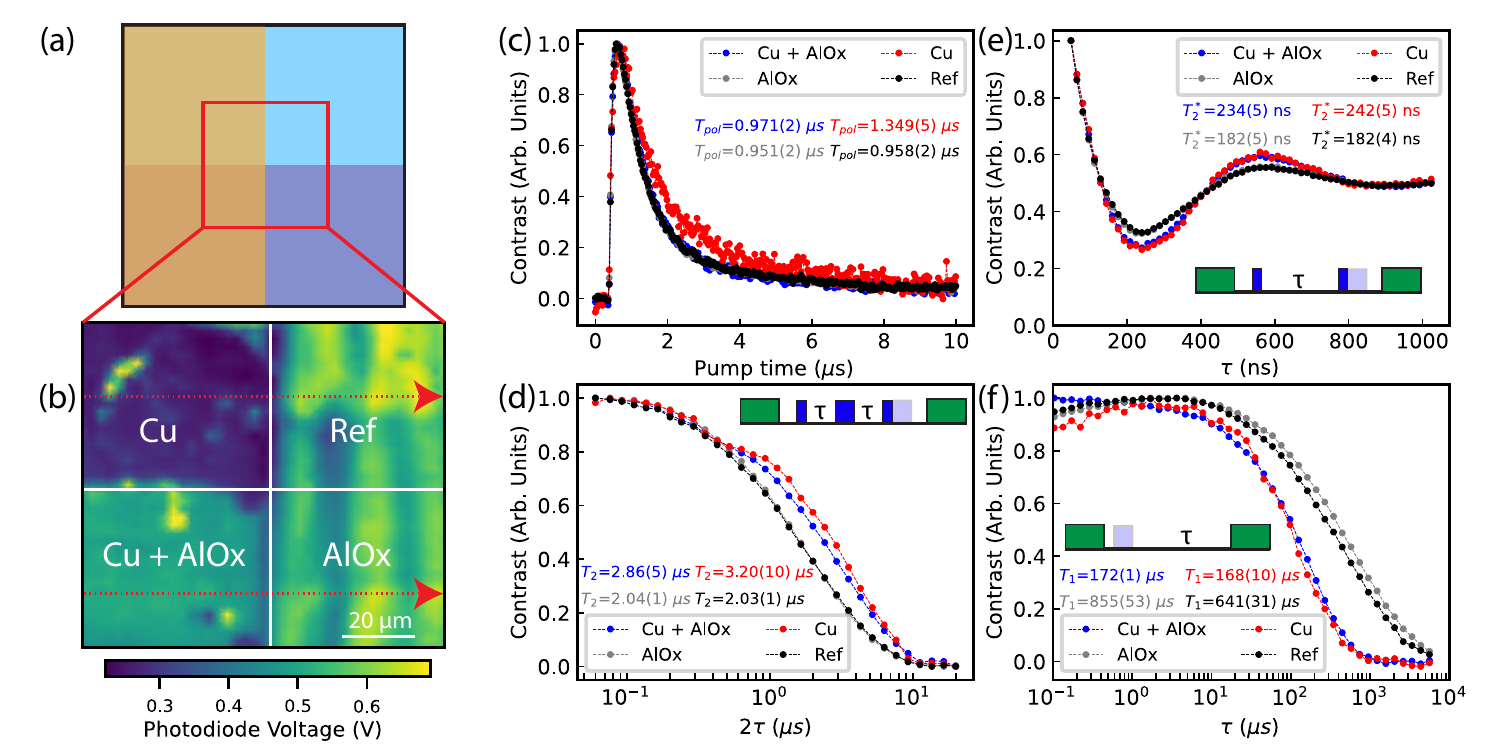}
\end{overpic}
\caption{\label{Fig1} (a) A schematic of the alumina and copper patterned on the diamond. (b) Photoluminescence(PL) image of the four quadrants of the 7 keV implanted diamond. The red dashed lines approximately indicate the lines where data was taken. (c-f) Normalized data from the four different regions for NV Polarization Time (c), Hahn Echo for $T_2$ (d), Ramsey interferometry for $T_2^*$ (e), and $T_1$ relaxation (f). The corresponding decay times of the different measurements are given as insets. For spin property measurements the pulse sequence used is shown.}
\end{figure}

We start with a series of electronic-grade diamond (Element 6) with natural \carbon{} abundance (1.1\%). The samples are implanted with \nitrogen{} and annealed to form NVs. The implant parameters of the samples are described in Table~\ref{ionEnergyFluence}. We implant these samples to achieve a $\mathrm{N_s}$ concentration of 100 ppm according SRIM. We follow the annealing procedure and oxidation treatment described in our previous work\cite{Henshaw_NQR}. After oxidation treatment, the samples have 2 nm of alumina ($\mathrm{Al_2O_3}$) deposited by atomic layer deposition (ALD) at 200C. The alumina is etched from half of the diamond and then a perpendicular half of the diamond has 50 nm of copper deposited using electron beam evaporation. This results in 4 regions: the bare diamond `Ref' region, diamond with alumina, `AlOx,' diamond with copper, `Cu,' and diamond with alumina and copper, `Cu + AlOx' (Fig.~\ref{Fig1}(a)). This is done so we can compare the effects of each environment in the same measurement series.

\begin{table}
\caption{\label{ionEnergyFluence} The energies, fluences, and SRIM-estimated depths for diamonds used in this work. All implants and SRIM simulations are done with an 8\textdegree{} tilt. Fluences are chosen so that the peak substitutional nitrogen ($\mathrm{N_s}$) concentration is 100 ppm. SRIM depth refers to the mode of the ion distribution.}

\begin{tabular}{lcr}
\hline
Energy (keV) & Fluence (ions/cm$^2$) &SRIM depth (nm)\\
\hline
3 & 1$\times10^{13}$ & 5.2\\
4 & 1.2$\times10^{13}$ & 6.6\\
5.5 & 1.6$\times10^{13}$  & 8.7\\
7 & 2$\times10^{13}$  & 10.6\\
15 & 3.25$\times10^{13}$  & 20.8\\
25 & 4.5$\times10^{13}$  & 33.5\\
\hline
\end{tabular}

\end{table}
We excite NV centers using a 532 nm laser with an optical power of 280 mW (before the objective) and focused down to a roughly 40 $\mathrm{\mu m}$ spot \cite{Henshaw_NQR}. The excitation provides a mechanism to spin initialize into the $\ket{0}$ spin state and readout the spin state from the spin-dependent fluorescence rate \cite{NVPhotophysics}. The samples rest on a sapphire substrate with a copper loop fabricated on it. The copper loop is connected to an amplified and gated microwave (mw) source to provide MW pulses for spin control. The photoluminescence (PL) is filtered using a 550 nm dichroic mirror, a 650 nm long-pass filter, and a 532 nm notch filter to suppress laser leakage and $\mathrm{NV^0}$ PL. The PL is detected by an \break A-CUBE-S1500-3 APD with variable gain. We perform scanning PL measurements using the stepper motors of a Thorlabs NanoMax 300.

\subsection{Sample characterization}
For each sample, we measure the PL intensity (Fig.~\ref{Fig1}(b)), NV polarization time (Fig.~\ref{Fig1}(c)), $T_2$ (Fig.~\ref{Fig1}(d)), $T_2^*$ (Fig.~\ref{Fig1}(e)), and $T_1$ (Fig.~\ref{Fig1}(f)). Representative data for a single point in each region is shown in Fig~\ref{Fig1}. The NV polarization time is measured by measuring two consecutive PL time traces, one with a mw $\pi$ pulse, and one without, and taking the difference between the time traces. This can be a probe of processes like Fluorescence Resonant Energy Transfer (FRET) \cite{NV_FRET}, as the reduced excited-state lifetime reduces polarization efficiency. We measure $T_2$ and $T_2^*$ using Hahn echo (Fig.~\ref{Fig1}(d)) and Ramsey interferometry (Fig.~\ref{Fig1}(e)) respectively. Both measurements are done on the $\ket{0} \leftrightarrow \ket{-1}$ transition of the NV ground-state spin sublevels and are thus not immune to strain and electric field fluctuations\cite{bauch_NvsT2,NV_Ultralong_Dephasing}. The second $\pi/2$ of both of these measurements toggles between a $\pi/2$ and $3\pi/2$ pulse and the difference between two consecutive measurements is taken to suppress noise. $T_2^*$ is the limiting timescale for ODMR linewidth and an important value for DC magnetometry with NVs. $T_2$ is the lifetime of a coherent state and is the limiting timescale for nanoscale NMR spectroscopy with NVs or sensing of low-frequency (100s kHz - 10s MHz) noise.  The $T_1$ measurement is referenced by applying a $\pi$ pulse on every other measurement and subtracting two consecutive measurements. The relaxation time, $T_1$, is sensitive to noise near the NV spin resonance frequency, providing a sensing mechanism over a wide range of frequencies tuned with a magnetic field\cite{T1_Relaxometry_Overview,T1_Relaxometry_NMR,widefieldT1EPR}.

In order to account for sample and microwave driving inhomogeneities, spin measurements are performed at points spaced by 25-50 $\mathrm{\mu m}$ along a pair of horizontal lines 500 $\mathrm{\mu m}$ (750 $\mathrm{\mu m}$ for $T_1$) long, from the 1Cu + AlOx' region into the 1AlOx' region and from the `Cu' to `Ref' regions (Red lines in Fig~\ref{Fig1}(b)). At each point we measure Rabi oscillations and NV resonance frequency along with the spin properties of the NVs. The PL rate, NV polarization time, contrast-weighted shot-noise ($C\sqrt{I_{pl}t_{ro}}$), $T_2^*$, $T_2$, $T_1$ are then averaged in each of the regions. No notable spatial dependence was observed within any given region or near the transition from one region to another for all properties except $T_1$. The $T_1$ of shallower NV ensembles did not reach a stable reference level until the probed position was a few hundred $\mathrm{\mu m}$ away from the metal edge. These measurements and processing steps are performed for all four regions across all six samples, with the exception of the 3 keV sample, where only the `Cu + AlOx' and `AlOx' areas are measured due to the signal-to-noise ratio of the `Cu' area being too small to achieve usable signal in a reasonable time. For the NV ensemble samples with depths less than 10 nm, no notable differences were observed between the `Ref' region and the 1AlOx' region. For the deepest two NV ensembles, a slight increase in PL was observed for 1AlOx' relative to `Ref.'

Due to the wide range of PL rates across the various NV ensembles, the gain of our APD needed to be adjusted from one diamond to another. Due to this, we do not quote explicit photon count rates as the APD responsivity and noise floor is not the same from diamond to diamond. To this end, all measurements or calculations that require a PL rate are expressed as ratios between regions. This still provides the critical information of relative PL rates in different regions across a single diamond.

\section{Results and Discussion}
\subsection{NV characterization}

\begin{figure}
\begin{overpic}{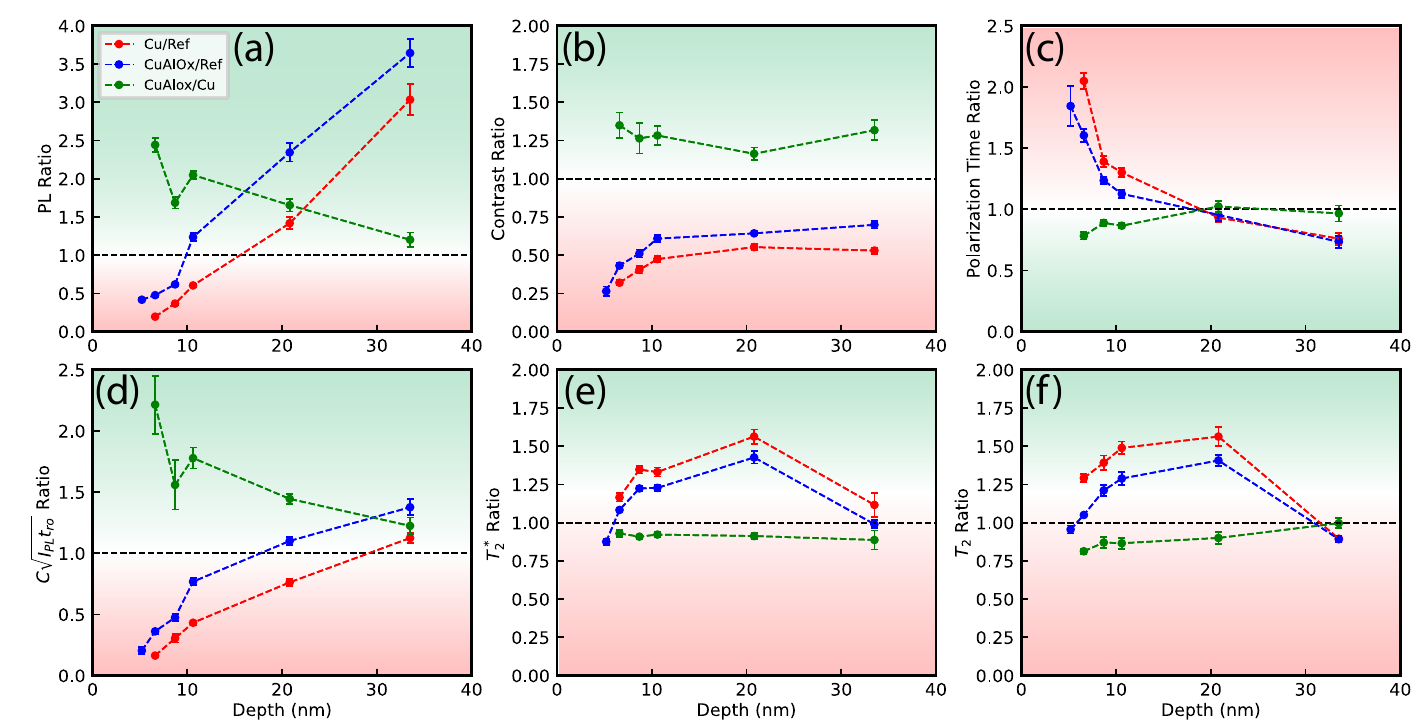}
\end{overpic}
\caption{\label{Fig2} Depth dependence of NV properties critical to sensing. The ratios between different regions for Photluminescence (PL) (a), Spin Contrast (b), Polarization time (c), contrast-weighted shot-noise($C\sqrt{I_{pl}t_{ro}}$) (d), $T_2^*$ (e), and $T_2$ (f), as a function of depth. The color gradient indicates sensitivity improvement (green) or worsening (red).Error bars are the greater of the fit errors or the standard deviation of the line cut data for a region.}
\end{figure}

We perform the previously mentioned measurements across a series of diamond samples with variable depth (Table ~\ref{ionEnergyFluence}). In Fig.~\ref{Fig2}, we show the ratios of NV properties critical to DC and AC sensing between different regions. We emphasize that, although for shallower NV ensembles the properties in both `Cu' and `Cu + AlOx' regions are strictly worse than the `Ref' region, the relevant comparison for sensing of nanoscale systems is `Cu + AlOx' to `Cu.' However, comparisons between the Cu coated regions and the `Ref' region do provide interesting insight into the changes in the environments in the differing regions caused by the integration of the material. 

The PL rate, shown as a function of depth for the different regions, in Fig.~\ref{Fig2}(a), shows a gradual increase in PL relative to the `Ref' region for both `Cu' and `Cu+AlOx' regions. Importantly, we saw the `Cu + AlOx' region has consistently high PL than the `Cu' region. The level of improvement decreases for deeper NV ensembles. For deeper NV ensembles we find the PL actually increases by as much as a factor of 4. This may be due to the metal and alumina increasing collection efficiency by acting as a mirror and reflective coating or an increase in the spontaneous emission rate via plasmonic interaction \cite{NV_Silver_Plasmonics}. 

We also observe the PL contrast between spin states in the `Cu + AlOx' and `Cu' regions to be much lower than that of the `Ref' region, plateauing at 10 nm. This may be due to background PL from $\mathrm{NV^0}$ for shallower NVs. Also, a slightly higher Rabi frequency was observed for the regions "Ref' and `AlOx' due to closer proximity to the mw loop. This could result in slightly lower contrast due to lower excitation bandwidth. When comparing `Cu+ AlOx' to `Cu,' we find an average 25\% increase in contrast with no clear depth dependence. 

We observe the polarization time increasing by as much as a factor of two for the shallowest NV ensembles (Fig.~\ref{Fig2}(c)), when compared to the `Ref' region. A fast polarization time for the NV is essential to reducing the overhead for of measurements. We attribute this increase to a reduced excited state lifetime due non-radiative relaxation cause by the metal through processes like FRET or Surface Energy Transfer (SET)\cite{NV_FRET}. For deeper ensembles, we find the polarization rate improves. We attribute this to the thin metal film acting as a  mirror and providing better laser excitation. The `Cu + AlOx' region is less impacted by the these processes due to an additional 2 nm stand off from the material.

An important parameter that appears in all shot-noise-limited sensitivity estimates is $C\sqrt{I_{pl}t_{ro}}$. This value is the fluorescence contrast between spin states times the shot-noise of a single measurement. When considering the comparison to the `Ref' region, we notice a dramatic drop, as low as a fifth the reference value (Fig.~\ref{Fig2}(d)). This is a mixture of notably reduced PL caused by band bending ionizing $\mathrm{NV^-}$, as well on non-radiative relaxation reducing photon generation from $\mathrm{NV^-}$\cite{NV_FRET} However, the comparison between `Cu + AlOx' and `Cu' regions sees a notable increase in this parameter for the shallowest NVs.

\subsection{Dephasing and Decoherence}
\begin{figure}
\begin{overpic}{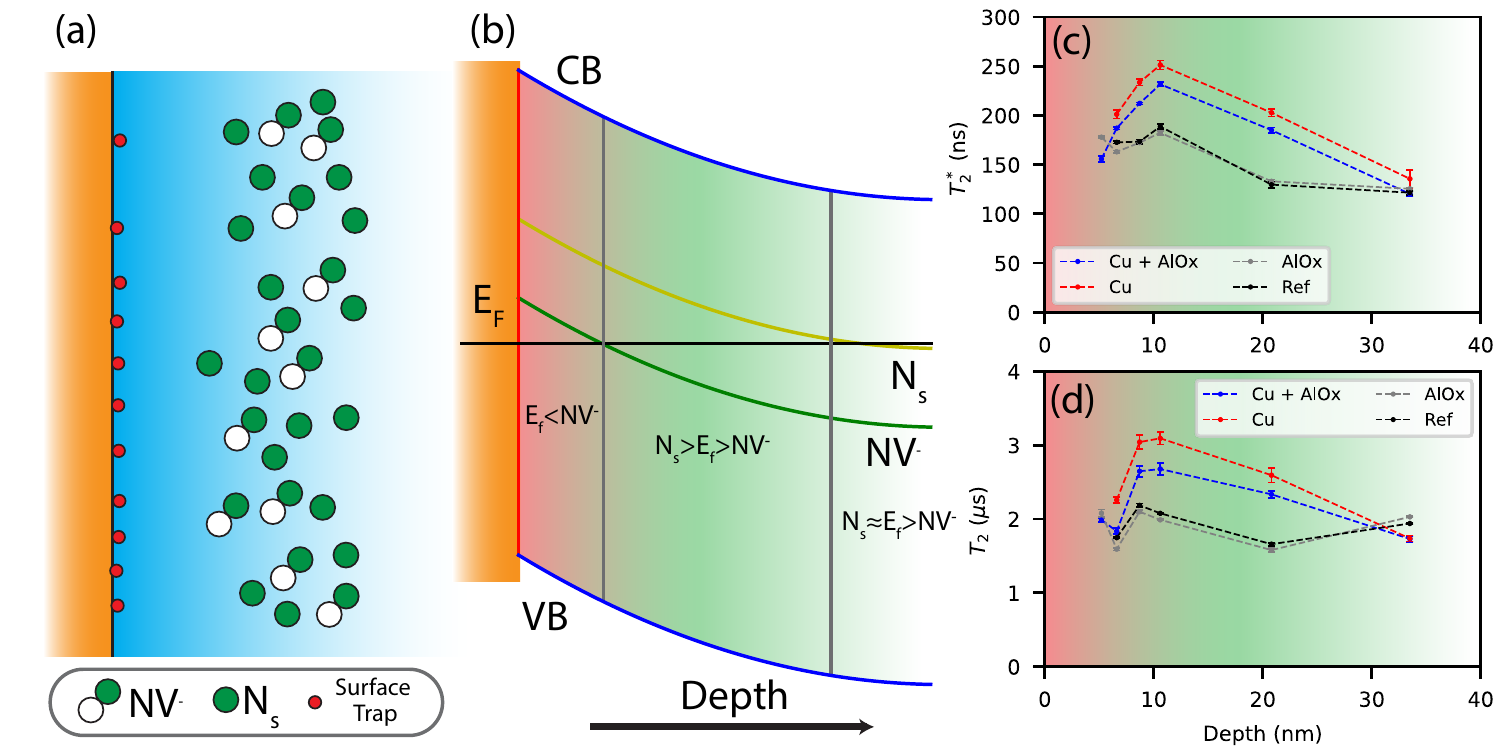}
\end{overpic}
\caption{\label{Fig3} Proposed mechanism behind improved coherence times (a) Diagram of the diamond surface indicating major decoherence dephasing sources like surface states and $\mathrm{N_s}$. (b) Band diagram depicting the NV and ($\mathrm{N_s}$) level under the influence of the band bending caused by the metal. (c) $T_2^*$ and (d) $T_2$ as a function of depth for the various regions.}
\end{figure}

We observe an interesting phenomenon when we look at $T_2^*$, the dephasing time (Fig.~\ref{Fig2}(e)), and $T_2$, the decoherence time (Fig.~\ref{Fig2}(f)). We find the coherence properties of the NVs improve under the metal. To explore why this happens we consider what are the major causes of decoherence and dephasing at this depth and nitrogen density. At the depths of our ensembles, surface noise from dangling bonds or other surface imperfections (Fig \ref{Fig3}(a)) has been seen to play a major role in decoherence\cite{depth_decoherence,surface_noise}. However, due to our very high nitrogen concentration (100 ppm), we posit paramagnetic noise from the $\mathrm{N_s}$ (Fig \ref{Fig3}(a)) is the dominant decoherence and dephasing source\cite{bauch_NvsT2}. 

We provide a qualitative explanation for the trend in $T_2^*$ and $T_2$ through a competition between these two noise sources, with the $\mathrm{N_s}$ being ionized by the band bending caused by the metal. The $\mathrm{N_s}$ is known to have a donor level 1.7 eV below the conduction band\cite{Ns_donor}. The $\mathrm{NV^-}$ ground state level has been found to be 2.6 eV below the conduction band\cite{NV_energy_level}. We propose that for very shallow NV centers, both the $\mathrm{N_s}$ and the NV are ionized by the band bending; the Fermi level drops below the defect levels(Fig \ref{Fig3}(b)). This regime provides a substantial decrease in PL, with spin properties dominated by surface noise, but also reducing the noise environment created by the $\mathrm{N_s}$. As the NVs get deeper, an ideal depth appears where there is sufficient $\mathrm{N_s}$ to charge NVs into the negatively charged state but not so much $\mathrm{N_s}$, that the NVs are still dominated by their noise. This regime is defined by the Fermi level being greater than the NV level but less than the $\mathrm{N_s}$ level. As NVs get sufficiently deep, the influence of the band bending becomes negligible, and the $\mathrm{N_s}$ keep their electron and the NVs become dominated by the nitrogen noise again, where the Fermi level approaches its bulk value determined by the nitrogen doping level. We show the explicit $T_2^*$ and $T_2$ values as a function of depth (Fig \ref{Fig3}(c,d)) with relevant regions highlighted according to our qualitative description.

An important note to this discussion is that this is an indirect effect. The metal or material is not directly reducing the $T_2$ or $T_2^*$; it is engineering the electronic environment in such a way that has an impact on the NV magnetic noise environment. Additionally, these changes are very much nitrogen concentration dependent; lower nitrogen densities may simply not have enough electrons to compensate the surface charge induced by the metal. Any sort of relaxometry using $T_2$ or $T_2^*$ would need to account for these environmental changes. For other modes of sensing, such as DC sensing which is $T_2^*$ limited\cite{taylor_sensitivity}, or nanoscale NMR, which is $T_2$ limited\cite{Henshaw_NQR,taylor_sensitivity},  this change in lifetimes is a strict benefit.
\subsection{Spin-Lattice Relaxation}
While $T_2$ and $T_2^*$ are critical parameters for sensing of DC magnetic fields and low frequency sources on the order of MHz, many proposals for using NVs to probe nanoscale electronic states use $T_1$ relaxometry\cite{demler_2d_SC,yacoby_review}. $T_1$ relaxometry has already been used as a probe of conductivity in metals\cite{NV_Conductivity,Kolkowitz_conductivity}. We use the previously-established techniques to demonstrate two features: we can recover the same information from $T_1$ relaxometry measurements with and without the alumina film and, for sensing Johnson noise in conductors, the additional stand-off provides no major deficits and even improves the sensitivity of $T_1$ relaxometry in the sample dominated regime, where the induced relaxation rate from the metal is much greater than the intrinsic relaxation rate.
\begin{figure}
\centering
\begin{overpic}{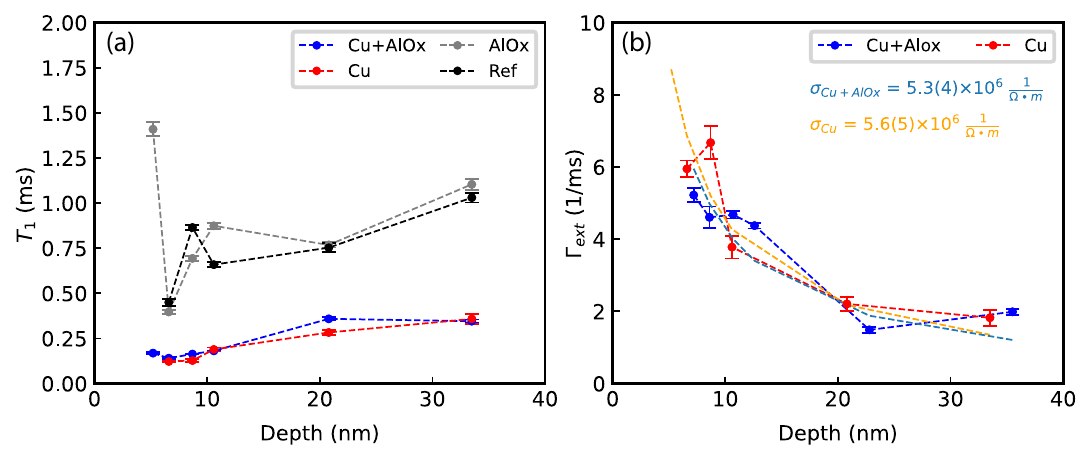}
\end{overpic}
\caption{\label{Fig4} Measurements of thin film conductivity. (a) $T_1$ for the various regions as a function of depth. (b)Relaxation rate from external sources, extracted from data in (a) as a function of depth. This is fit to Eqn. ~\ref{conductivity} to estimate the conductivity of the thin metal film. The extracted conductivities agree with each other and previous measurements of thin copper films.}
\end{figure}
In order extract the conductivity from our $T_1$ data, shown in Fig~\ref{Fig4}(a), we follow the process laid out in Ref.\cite{NV_Conductivity}. We must determine the relaxation rate from external sources, $\Gamma_{ext}(d,\sigma)$. Here d is the stand-off between the NV ensemble and the metal, and $\sigma$ is the conductivity of the metal. We do this by measuring the intrinsic relaxation rate, $\Gamma_{NV,int}$, of NVs unperturbed by the metal and the relaxation rate of NVs affected by the metal, $\Gamma_{NV}(d,\sigma)$. Due to the nature of our experimental configuration, we can use the `Ref' region to determine our $\Gamma_{NV,int}$. With this we can use the following equation to determine the extrinsic relaxation rate, $\Gamma_{ext}$
\begin{equation}\label{relaxation}
    \Gamma_{NV}(d,\sigma) = \Gamma_{ext}(d,\sigma) + \Gamma_{NV,int}.
\end{equation}
With $\Gamma_{ext}$, we can fit the depth dependence to determine the conductivity to the following function
\begin{equation}\label{conductivity}
    \Gamma_{ext}(d,\sigma) = \gamma_{e}^2\frac{\mu_0^2k_BT\sigma}{8\pi}\frac{1}{d},
\end{equation}
where $\gamma_e$ is the electron gyromagentic ratio, $\mu_0$ is the vacuum magnetic permeability, $k_B$ is the Boltzmann constant, and $T$ is the temperature. $\sigma$ is left as a free parameter to vary for the fit. Fitting was attempted to account for the film thickness, but the thickness was consistently fit to an arbitrarily large value, indicating our range of depths is much smaller than the film thickness.

The results of the calculation of $\Gamma_{ext}$ and the estimation of the conductivity are shown Fig~\ref{Fig4}(b). A note, regarding the data for the `Cu + AlOx' region, the x-axis is shifted by 2 nm to account for the additional 2 nm spacing provided by the alumina. The most important feature is that the the determined conductivities agree with each quite well, and both values agree with previous examinations of thin copper films\cite{thin_film_conductivity}. 

It is worth noting that $T_1$ of the `Cu + AlOx' region did not differ from the `Cu' region substantially (Fig~\ref{Fig4}(a)). This is because, for the case of Johnson noise, the induced relaxation rate scales as $d^{-1}$ (see Eqn.~\ref{conductivity}). In this case, an additional 2 nm is not a substantial change within the error of our measurement.

\subsection{Influence on Sensitivity}
We now discuss the idea of sensitivity and the impact that the AlOx layer has on the sensitivity of the NV ensembles for different sensing modalities. The sensitivity is the noise floor of a measurement given 1 second of integration time. This definition means a lower sensitivity provides a better sensor. We compare the sensitivity of NV ensembles in the different regions at different depths. As mentioned earlier, the contrast-weighted shot-noise appears in all sensitivity estimates with $\eta \propto \frac{1}{C\sqrt{I_{pl}t_{ro}}}$, where $\eta$ is used to denote sensitivity. Due to how dramatic the changes in the contrast-weighted shot-noise are from region to region, it plays a dominant role when comparing different regions' sensitivity. This can be seen by looking at the dependencies of the sensitivity in different modes:
\begin{equation}\label{DC_sens}
    \eta_{DC} \propto \frac{1}{C\sqrt{I_{pl}t_{ro}}}\frac{1}{\sqrt{T_2^*}},
\end{equation}
\begin{equation}\label{AC_sens}
    \eta_{AC}^{Mean} \propto \frac{1}{C\sqrt{I_{pl}t_{ro}}}\frac{1}{\sqrt{T_2}},
\end{equation}
\begin{equation}\label{AC_sens_var}
    \eta_{AC}^{Var} \propto \frac{1}{C\sqrt{I_{pl}t_{ro}}}\frac{1}{T_2^{3/2}}.
\end{equation}
We observe improvements in the $T_2$ and $T_2^*$ by a factor of 1.5 when compared to the `Ref' region. This results in improvements of the sensitivity by a factor of roughly 1.2 ($\sqrt{1.5}$) for $\eta_{DC}$ and $\eta_{AC}^{Mean}$, the sensitivities relevant for DC magnetometry and AC magnetometry. For nanoscale NMR measurements, the sensitivity, $\eta_{AC}^{Var}$, will improve by a factor of 1.8 ($1.5^{3/2}$), due to the stronger dependence on $T_2$ for variance detection. While these are sizable improvements, the constrast weighted shot-noise is reduced by a factor of 6 as compared to the reference. In this regard, the sensitivity, as compared to the reference, is strictly worse except for the deeper ensembles which see a slight improvement.

\begin{figure}
\centering
\begin{overpic}{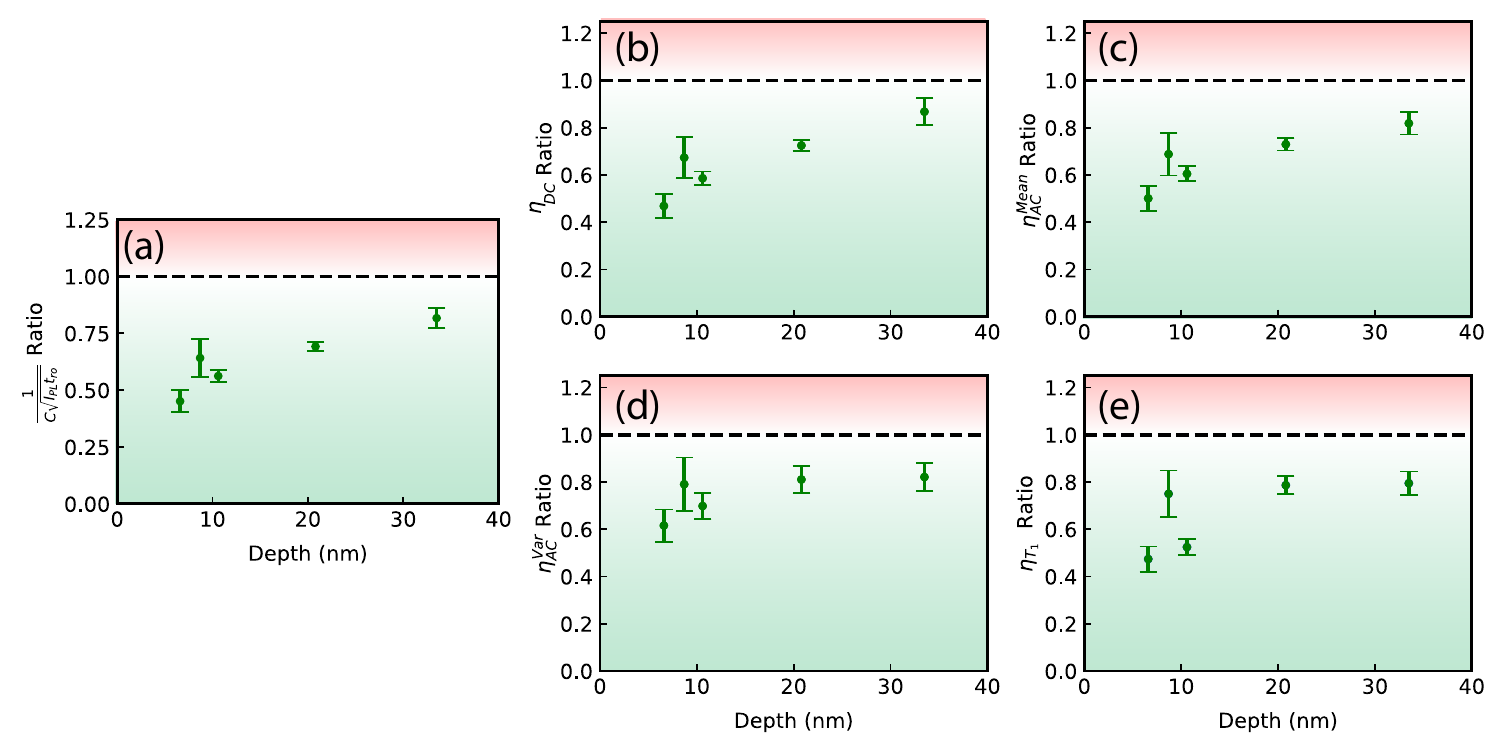}
\end{overpic}
\caption{\label{Fig5} Sensitivity of NVs near thin metal films in different sensing modes.(a) The inverse contrast-weighted shot-noise ratios as a function of depth for the `Cu+AlOx' region and the `Cu' region. $\eta_{DC}$(b), $\eta_{AC}^{Mean}$(c), $\eta_{AC}^{Var}$(d), $\eta_{T_1}$(e), ratios as function of depth for the `Cu+AlOx' region and the `Cu' region.}
\end{figure}

Our main focus is the comparison between the two copper coated regions. We do note that the sensitivity ratios when compared to the reference region reaches 1 at around 10 nm deep and goes below 1, meaning improved sensitivity, for deeper ensembles. This change is can be appreciable, as much as 0.5 and the causes for this increase are the increased PL rate and $T_2$ or $T_2^*$. When the `Cu+AlOx' region is compared to the `Cu' region, the relevant comparison for sensing of quantum materials, it is clearly more sensitive. Although the insulating layer does decrease the degree of improvement for $T_2$ and $T_2^*$,it improves the contrast-weighted shot-noise (inverse shown in Fig ~\ref{Fig5}(a), by as much as a factor of 2, resulting in a reduction (improvement) in sensitivity of a factor of 2. There is also the extreme case of the 3 keV implanted sample where measurement in the `Cu' area were not feasible due to very low signal-to-noise ratio. 

For all forms of quantum sensing, the `Cu+AlOx' region improves the sensitivity for all measured NV ensemble depths when compared to the `Cu' region. The impact is most prominent for the shallowest NV ensembles, but still a 20-40\% improvement for the deeper NV ensembles. As mentioned, the changes in PL rate dominate resulting in very similar looking data for all modes of sensing. For DC sensing, the relative change in $T_2^*$ was effectively flat (see Fig.~\ref{Fig2}(e)). The DC and AC mean sensitivities scale as the square root of the $T_2^*$ and $T_2$ respectively, further flattening the small difference between the two regions resulting in trends dominated by the change in $C\sqrt{I_{pl}t_{ro}}$ (Fig.~\ref{Fig5}(b,c)). A slight deviation from this trend is observed in the AC variance sensitivity, which scales like $T_2^{3/2}$. In Fig\ref{Fig2}(f),we saw a slight depth dependence in $T_2$, showing the $T_2$ ratio between the regions under discussion being less than 1 for shallow ensembles. The stronger $T_2$ dependence in variance sensing amplifies this dependence and a slightly weaker improvement in $\eta_{AC}^{Var}$ as a function of depth is observed.

The sensitivity of $T_1$ relaxometry needs to be treated separately due to the fact that the measurement revolves around observing changes in $T_1$. In the regime where $\Gamma_{ext} \gg \Gamma_{int}$, the sensitivity can be estimated\cite{T1_Relaxometry_Overview,T1_Relaxometry_NMR} by
\begin{equation}\label{T1_sens}
    \eta_{T_1} \propto \frac{1}{C\sqrt{I_{pl}t_{ro}}}\frac{1}{\sqrt{\Gamma_{ext}}},
\end{equation}
This is a sensitivity to changes in $T_1$ with respect to the intrinsic $T_1$. If $\Gamma_{ext} = 1/T_1$, this means a shorter $T_1$ takes less time to sense. The sensitivity ratio between the `Cu + AlOx' and `Cu' regions is shown in Fig.~\ref{Fig5}(e) for different depths. As previously mentioned, the alumina provided minimal change on the $T_1$, thus this sensitivity is dominated by the PL improvement provided by the insulating layer. For the shallowest samples we see sensitivity improvements as much as a factor of 2.
\section{Conclusions}
This work highlights the importance of how integrating a material on the diamond surface can impact NV performance. Though we chose thin copper films, hardly a low dimensional quantum material, to demonstrate these affects, we view this as a sort of worst-case scenario. Not all process observed here will be observed to the same degree in other materials. Low dimensional materials, like magic-angle graphene, will not only provide a magnetic noise source as it approaches superconductivity\cite{demler_2d_SC}, it will provide an acceptor for processes like FRET\cite{NV_FRET} and can reduce PL intensity. Surface charging caused by the integration of the material may cause band bending, making changes in spin properties difficult to isolate due to changes in the magnetic environment. Our approach of incorporating an insulating layer of alumina, has reduced the impact of the integration process and enabled the use of shallower NV ensembles for nanoscale quantum sensing.

We have characterized the PL rate, NV polarization time, $T_2$, $T_2^*$, and $T_1$ for a series of NV ensembles at variable depths. These measurements were performed in 4 different regions: with copper in direct contact with the diamond surface, `Cu', with copper insulated from the diamond surface by 2 nm of alumina, `Cu+ AlOx', just the alumina layer, `AlOx', and bare diamond, `Ref.' We observed a general decrease in PL rate for NV ensembles closer to the copper film, that could be tuned by with the addition of an insulating layer. We also found a relative increase of the NV polarization time as NV ensembles approached the film. We saw a non-monotonic improvement in $T_2$ and $T_2^*$ over the characterized depths. We attribute this improvement to band bending caused by the metal film, ionizing paramagnetic noise sources near the NV center inside the diamond.

We considered the impact of the alumina film in terms of sensing for different sensing modalities. For the deepest NV ensembles, the alumina and metal seemed to provide an overall improvement, in terms of sensitivity when compared to the reference region. In general, NVs at or deeper than 10 nm had sensitivities on par with the reference reference region, though the exact contributions to that sensitivity was different that that of the reference. For sensing with $T_1$ relaxometry, the `Cu + AlOx' region provides consistently superior sensitivity to the `Cu' region providing as much as a factor of 2 increase in sensitivity, decrease of integration time by a factor of 4, for $T_1$ relaxometry measurements.

There are further techniques that could be developed to suppress the influence of integrated materials. Recent work on using applied electric fields to engineer the charge environment in a more deterministic means has been demonstrated for single NVs\cite{NV_charge_engineering}. Another approach would be to use another donor to provide charge such as phosphorus. Such co-doping has been shown to provide high NV conversion efficiency, provide, better NV properties, and would provide more charge to passivate the induced charge with minimal cost to sensor quality\cite{charge_engineering_1,Charge_engineering_2,NV_Phosphorus_doped}

\section*{Acknowledgements}

Sandia National Laboratories is a multi-mission laboratory managed and operated by National Technology and Engineering Solutions of Sandia, LLC, a wholly owned subsidiary of Honeywell International, Inc., for the DOE's National Nuclear Security Administration under contract DE-NA0003525. This work was funded, in part, by the Laboratory Directed Research and Development Program and performed, in part, at the Center for Integrated Nanotechnologies, an Office of Science User Facility operated for the U.S. Department of Energy (DOE) Office of Science. This paper describes objective technical results and analysis. Any subjective views or opinions that might be expressed in the paper do not necessarily represent the views of the U.S. Department of Energy or the United States Government.

\section*{Data Availability}
The data that support the findings of this study are available from the corresponding author upon reasonable request.

\newcommand{\newblock}{}
\bibliographystyle{unsrt}
\bibliography{metalBib}

\end{document}